\documentclass[prb,aps,twocolumn,showpacs,ams]{revtex4}
\usepackage{graphicx}
\usepackage{dcolumn}
\usepackage{latexsym}
\usepackage{color}

\begin{document}


\title{
Knight shift spectrum in vortex states 
in $s$-wave and $d$-wave superconductors \\ 
on the basis of Eilenberger theory}



\author{Kenta K. Tanaka} 
\email[]{ktanaka@mp.okayama-u.ac.jp}
\affiliation{
Department of Physics, Okayama University, Okayama 700-8530, JAPAN}

\author{Masanori Ichioka}
\email[]{ichioka@cc.okayama-u.ac.jp}
\affiliation{
Department of Physics, Okayama University, Okayama 700-8530, JAPAN}

\author{Noriyuki Nakai} 
\affiliation{
Department of Physics, Okayama University, Okayama 700-8530, JAPAN}

\author{Kazushige Machida} 
\affiliation{
Department of Physics, Okayama University, Okayama 700-8530, JAPAN}


\date{\today}

\begin{abstract}
From the spatial structure of vortex lattice state calculated 
by Eilenberger theory, we study the resonance line shape of 
Knight shift of the paramagnetic moments in the $s$-wave and 
the $d$-wave superconductors, comparing with 
the Redfield pattern of the internal field distribution. 
We discuss the deviation from the temperature dependence of 
the Yosida function, and the magnetic field dependence of 
the paramagnetic susceptibility. 
In addition to the calculation in the clean limit,  
influences of the impurity scattering are estimated in the Born limit 
and in the unitary limit. 
These results are helpful for the analysis 
of NMR experiments to know properties of the superconductors. 
\end{abstract}



\maketitle

\section{Introduction}

In the study of superconductivity,  
the observation of Knight shift by NMR experiments is 
an important method to identify the pairing symmetry. 
The Knight shift is related to the paramagnetic susceptibility, 
and it is suppressed below the superconducting transition temperature,  
if the superconductivity is the spin-singlet pairing.~\cite{Yosida,Tou}   
At a zero field, the temperature ($T$) dependence of the Knight shift 
is described by the Yosida function.~\cite{Yosida} 
It shows either an exponential $T$-dependence at low $T$ 
in the $s$-wave superconductors with the full gap, 
or a power-law $T$-dependence in anisotropic superconductors with nodes. 
On the other hand, the paramagnetic susceptibility $\chi$ is 
proportional to the electronic specific heat at low $T$, 
since both quantities are proportional to zero-energy density of states (DOS). 
In the $s$-wave pairing, we expect the linear $H$-dependence of $\chi$ 
at low $H$ and low $T$ region.~\cite{IchiokaPara,IchiokaIntech} 
In the $d$-wave pairing with line nodes, 
we expect the relation $\chi \propto \sqrt{H}$ due to the Volovik 
effect.~\cite{Volovik,IchiokaPara,IchiokaIntech,Zheng,Koutroulakis} 
Therefore, by the careful observations of the $T$- and $H$-dependence 
of the Knight shift, we may obtain valuable information to identify 
the pairing symmetry of the superconductivity.   
However, the NMR experiment to detect the Knight shift 
is usually performed in the vortex states under static magnetic fields. 
Therefore, in order to correctly analyze the Knight shift, 
we have to evaluate properties of the resonance line shape 
of the NMR spectrum considering   
the non-uniform spatial structure of paramagnetic moments 
in the vortex states.

In the NMR experiment, the spectrum of the nuclear spin 
resonance is determined by the internal magnetic field 
and the hyperfine coupling to the spin of the conduction electrons. 
Therefore, in a simple consideration, the effective field 
for the nuclear spin is given by 
$B_{\rm eff}({\bf r})=B({\bf r})+ A_{\rm hf}M_{\rm para}({\bf r})
$,~\cite{IchiokaFFLO,IchiokaIntech,SuzukiFFLO,Kumagai2011} 
where $B({\bf r})$ is the internal field distribution, 
$M_{\rm para}({\bf r})$ is the paramagnetic moment of conduction electrons,  
and $A_{\rm hf}$ is a hyperfine coupling constant depending 
on species of the nuclear spins. 
The resonance line shape of NMR is given by 
\begin{eqnarray} 
P(\omega)=\int \delta(\omega-B_{\rm eff}({\bf r}) ){\rm d}{\bf r}, 
\end{eqnarray} 
i.e., the intensity at each resonance frequency $\omega$ comes from 
the volume satisfying $\omega=B_{\rm eff}({\bf r})$ in a unit cell. 
When the contribution of the hyperfine coupling is dominant,  
the NMR signal selectively detects $M_{\rm para}({\bf r})$. 
This is the experiment observing the Knight shift.  
As the resonance line shape of the NMR spectrum for the Knight shift, 
we calculate the distribution function 
$P(M)=\int \delta(M-M_{\rm para}({\bf r}) ){\rm d}{\bf r}$
from the spatial structure of $M_{\rm para}({\bf r})$. 
On the other hand, in the case of negligible hyperfine coupling, 
the NMR signal is determined by $B({\bf r})$. 
This resonance line shape in the vortex lattice state 
is called ``Redfield pattern''.~\cite{Redfield,Kung,IchiokaD1}  
The resonance line shape is given by the distribution function 
$P(B)=\int \delta(B-B({\bf r}) ){\rm d}{\bf r}$ 
calculated from the internal field $B({\bf r})$.

Since the hyperfine coupling constant has 
different values for different nuclei, 
whether we observe the Redfield pattern of $P(B)$ or 
the Knight shift spectrum of $P(M)$ depends on the target nuclei 
in the NMR experiment, even in same superconductors.  
The distributions of $P(M)$ and $P(B)$ were sometimes confused 
in analysis of the NMR resonance line shape in the vortex states. 
Thus, it is important to clarify differences of the behaviors 
between $P(B)$ and $P(M)$. 

 The purpose of this work is to calculate the Knight shift spectrum $P(M)$ 
and the Redfield pattern $P(B)$ in the vortex lattice state on 
the basis of Eilenberger theory,~\cite{eilenberger,Klein1987,IchiokaD1,IchiokaPara} 
and discuss differences between them. 
We quantitatively estimate the $T$-dependence and the $H$-dependence 
of the Knight shift spectrum. 
We discuss their behaviors depending on the pairing symmetries, 
i.e., $s$-wave pairing and $d$-wave pairing. 
In addition to the clean limit, 
we study the influence of the impurity scatterings  
in the Born limit and the unitary limit, 
where the residual DOS appears 
in the superconducting 
state.\cite{kubert,hirschfeld1988,thuneberg,kato2000,eshrig2002,hayashi2002,Miranovic,hayashi2005,sauls2009}  
We discuss how the impurity scattering changes the NMR resonance line shape. 

This paper is organized as follows. 
After the introduction, 
formulation of our calculation is explained 
in Sec. \ref{sec:formulation}. 
In  Sec. \ref{sec:s},  
after calculating the spatial structure 
of $M_{\rm para}({\bf r})$ and $B({\bf r})$,  
we discuss the $T$- and $H$-dependences of 
the resonance line shape $P(M)$ and $P(B)$ 
in the clean limit and in the presence of non-magnetic impurity scatterings 
for the $s$-wave pairing. 
The results for the $d_{x^2-y^2}$-wave pairing are 
reported in Sec. \ref{sec:d}.  
The last section is devoted to summary. 


\section{Formulation by selfconsistent quasiclassical theory}
\label{sec:formulation}

We calculate the spatial structure of vortices in the vortex lattice state 
by quasiclassical Eilenberger 
theory,~\cite{eilenberger,Klein1987,IchiokaD1,IchiokaPara,IchiokaIntech} 
including impurity scatterings. 
\cite{thuneberg,kato2000,eshrig2002,hayashi2002,Miranovic,hayashi2005,sauls2009}
In order to estimate paramagnetic susceptibility, 
we include weak Zeeman term $\mu_{\rm B}B({\bf r})$, 
where 
$\mu_{\rm B}$ is a renormalized Bohr 
magneton.~\cite{WatanabeKita,klein2000,IchiokaFFLO,IchiokaPara,IchiokaIntech,SuzukiFFLO} 
The quasiclassical theory assumes that 
the atomic scale is enough small compared to 
the superconducting coherence length $\xi$, 
and we focus the spatial structure in the order of $\xi$-scale.  
The quasiclassical condition is satisfied in many superconductors. 
We also assume that the size of the impurity is in the atomic scale, 
so that the impurity does not work as a pinning center for vortices. 
Thus we consider the case of uniform vortex lattice points in this work. 
The impurity scatterings contribute to the self-energy of 
the electronic states.

To obtain quasi-classical Green's functions 
$g({\rm i}\omega_n,{\bf k},{\bf r})$, 
$f({\rm i}\omega_n,{\bf k},{\bf r})$ and 
$f^\dagger({\rm i}\omega_n,{\bf k},{\bf r})$, 
we solve Ricatti equation obtained from Eilenberger equations 
\begin{eqnarray} &&
\left\{ \omega_n +{\rm i}{\mu}B +\frac{1}{\tau} \langle g \rangle_{\bf k} 
+{\bf v} \cdot\left(\nabla+{\rm i}{\bf A} \right)\right\} f 
\nonumber \\ && \qquad 
=\left( \Delta \phi + \frac{1}{\tau} \langle f \rangle_{\bf k}  \right) g, 
\nonumber 
\\ && 
\left\{ \omega_n +{\rm i}{\mu}B + \frac{1}{\tau} \langle g \rangle_{\bf k} 
-{\bf v} \cdot\left( \nabla-{\rm i}{\bf A} \right)\right\} f^\dagger
\nonumber \\ && \qquad 
=\left( \Delta^\ast \phi^\ast 
+ \frac{1}{\tau} \langle f^\dagger \rangle_{\bf k}  \right) 
g  , \quad 
\label{eq:Eil}
\end{eqnarray} 
where $g=(1-ff^\dagger)^{1/2}$, 
${\mu}=\mu_{\rm B} B_0/\pi k_{\rm B}T_{\rm c}$, and 
${\bf v}={\bf v}_{\rm F}/v_{{\rm F}0}$ with Fermi velocity ${\bf v}_{\rm F}$ and 
$v_{\rm F 0}=\langle {\bf v}_{\rm F}^2 \rangle_{\bf k}^{1/2}$.  
$\langle \cdots \rangle_{\bf k}$ indicates the Fermi surface average. 
${\bf k}$  is the relative momentum of the Cooper pair 
on the Fermi surface, 
and ${\bf r}$ is the center-of-mass coordinate of the pair. 
In our calculations, length, temperature, Fermi velocity, 
magnetic field and vector potential are, respectively, measured in unit of 
$\xi_0$, $T_{\rm c}$, $v_{\rm F0}$, $B_0$ and $B_0 \xi_0$. 
Here, $\xi_0=\hbar v_{\rm F0}/2 \pi k_{\rm B} T_{\rm c}$, 
$B_0=\phi_0 /2 \pi \xi_0^2$ with the flux quantum $\phi_0$.  
$T_{\rm c}$ is superconducting transition temperature in the clean limit at 
a zero magnetic field.  
The energy $E$, pair potential $\Delta$ and Matsubara frequency $\omega_n$ 
are in unit of $\pi k_{\rm B} T_{\rm c}$. 

\begin{figure}
\begin{center}
\includegraphics[width=8.0cm]{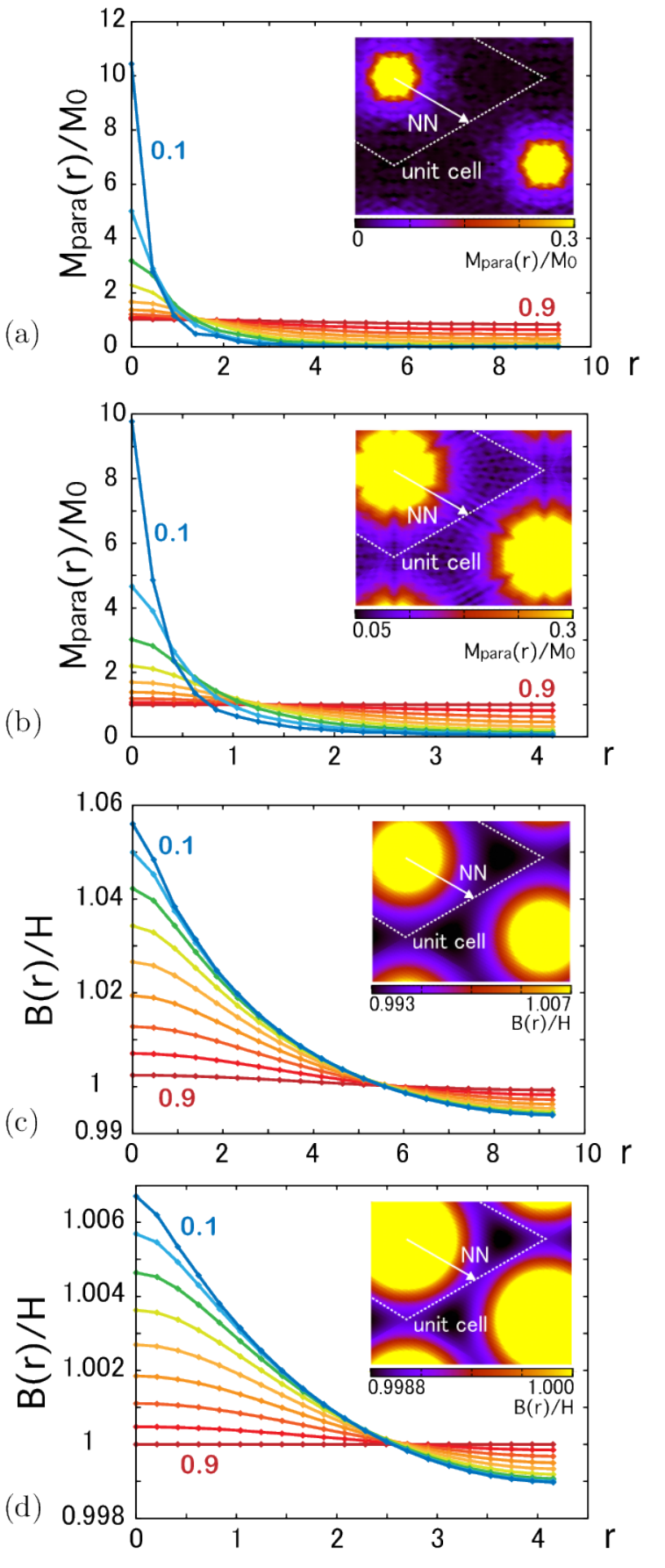}
\end{center}
\vspace{-0.7cm}
\caption{\label{fig1}
(Color online) 
(a) 
Profiles of the paramagnetic moment $M_{\rm para}({\bf r})$ at $H=0.02$ 
as a function of radius $r/\xi_0$ from the vortex center 
along the nearest neighbor (NN) directions 
at $T/T_{\rm c}=0.1, \ 0.2, \cdots, 0.9$.
The inset shows a density plot of spatial structure 
of $M_{\rm para}({\bf r})$ at $T/T_{\rm c}=0.1$. 
Peak height at the vortex core is truncated in the density plot. 
Dashed lines indicate a unit cell of the vortex lattice in our calculations. 
(b) The same as (a), but at $H=0.1$. 
(c) 
Profiles of the internal field distribution $B({\bf r})$ at $H=0.02$ 
as a function of $r/\xi_0$ at $T/T_{\rm c}=0.1, \ 0.2, \cdots, 0.9$. 
The inset shows a density plot of spatial structure of $B({\bf r})$. 
(d) The same as (c), but at $H=0.1$.  
These are for the $s$-wave pairing in the clean limit.  
}
\end{figure}

For simplicity, we consider the spin-singlet pairing 
on the two-dimensional cylindrical Fermi surface, 
${\bf k}=(k_x,k_y)=k_{\rm F}(\cos\theta_k,\sin\theta_k) $ and Fermi velocity 
${\bf v}_{\rm F}=v_{\rm F0} {\bf k}/k_{\rm F}$. 
The order parameter is 
$\tilde\Delta({\bf r},{\bf k})=\Delta({\bf r}) \phi({\bf k})$
with the pairing function 
$\phi({\bf k})=\sqrt{\mathstrut 2}(k_x^2-k_y^2)/k_{\rm F}^2$  
for the $d_{x^2-y^2}$-wave pairing,  
or $\phi({\bf k})=1$ for the $s$-wave pairing.
As magnetic fields are applied to the $z$ axis, 
the vector potential is given by   
${\bf A}({\bf r})=\frac{1}{2} {\bf H} \times {\bf r}
 + {\bf a}({\bf r})$ in the symmetric gauge, 
where ${\bf H}=(0,0,H)$ is a uniform flux density, 
and ${\bf a}({\bf r})$ is related to the internal field 
${\bf B}({\bf r})={\bf H}+\nabla\times {\bf a}({\bf r})$.
As shown in the insets of Fig. \ref{fig1}, 
the unit cell of the vortex lattice is given by 
${\bf r}=s_1({\bf u}_1-{\bf u}_2)+s_2{\bf u}_2$ with 
$-0.5 \le s_i \le 0.5$ ($i$=1, 2), ${\bf u}_1=(a_x,0,0)$,   
${\bf u}_2=(a_x/2,a_y,0)$ and $a_x a_y H=\phi_0$.  
$a_y/a_x=\sqrt{3}/2$ for the triangular vortex lattice, 
and $a_y/a_x=1/2$ for the square vortex lattice.

We consider the case of non-magnetic $s$-wave impurity scatterings  
with impurity strength $u_0$, 
and treat the self-energy by the $t$-matrix 
approximation.\cite{thuneberg,kato2000,eshrig2002,hayashi2002,Miranovic,hayashi2005,sauls2009} 
Thus, $1/\tau$ in Eq. (\ref{eq:Eil}) is given by  
\begin{eqnarray} 
\frac{1}{\tau}=\frac{1/\tau_0}{\cos^2\delta_0 
+ ( \langle g \rangle_{\bf k}^2 +\langle f \rangle_{\bf k}
  \langle f^\dagger \rangle_{\bf k} ) \sin^2 \delta_0}
\end{eqnarray} 
and $\delta_0=\tan^{-1}(\pi N_0 u_0)$. 
The scattering time $\tau_0$ in the normal state is given by 
$1/\tau_0 = n_s N_0 u_0^2 / (1 + \pi^2 N_0^2 u_0^2)$,  
where $n_s$ is the number density of impurities, 
and $N_0$ is the DOS at the Fermi energy in the normal state.
In this paper, we write 
$\hbar/ 2\pi k_{\rm B}T_{\rm c} \tau_0  \rightarrow 1/\tau_0$, 
since the scattering time $\tau_0$ is 
in unit of  $ 2\pi k_{\rm B}T_{\rm c} /\hbar$. 
The relation to the mean free path $l=v_{\rm F0} \tau_0$ 
and the zero-temperature coherence 
length $\xi=\Delta_0/\pi k_{\rm B} T_{\rm c}$ is given by 
$l/\xi= (2 \pi k_{\rm B}T_{\rm c} \tau_0/\hbar )( \Delta_0 / 2  k_{\rm B}T_{\rm c}) 
\rightarrow \tau_0 \Delta_0 / 2  k_{\rm B}T_{\rm c}  $ in our unit. 
In the Born limit of weak impurity scattering potential, 
$\delta_0 \rightarrow 0$. 
In the unitary limit of strong scattering potential, 
$\delta_0 \rightarrow \pi/2$.

As for selfconsistent conditions, 
the pair potential is calculated by the gap equation 
\begin{eqnarray}
\Delta({\bf r})
= g_0N_0 T \sum_{0 < \omega_n \le \omega_{\rm cut}} 
 \left\langle \phi^\ast({\bf k}) \left( 
    f +{f^\dagger}^\ast \right) \right\rangle_{\bf k} 
\label{eq:scD} 
\end{eqnarray} 
with 
$(g_0N_0)^{-1}=  \ln T +2 T
        \sum_{0 < \omega_n \le \omega_{\rm cut}}\omega_n^{-1} $. 
We use $\omega_{\rm cut}=20 k_{\rm B}T_{\rm c}$.
The vector potential for the internal magnetic field 
is selfconsistently determined by 
\begin{eqnarray}
\nabla\times \left( \nabla \times {\bf A} \right) 
=\nabla\times {\bf M}_{\rm para}({\bf r})
-\frac{2T}{{{\kappa}}^2}  \sum_{0 < \omega_n} 
 \left\langle {\bf v}_{\rm F} 
         {\rm Im} g  
 \right\rangle_{\bf k}, 
\qquad
\label{eq:scH} 
\end{eqnarray} 
where 
${\bf M}_{\rm para}({\bf r})=(0,0,M_{\rm para}({\bf r}))$ 
with 
\begin{eqnarray}
M_{\rm para}({\bf r})
=M_0 \left( 
\frac{B({\bf r})}{H} 
- \frac{2T}{{\mu} H }  
\sum_{0 < \omega_n}  \left\langle {\rm Im} \left\{ g \right\} 
 \right\rangle_{\bf k}
\right) ,
\label{eq:scM} 
\end{eqnarray} 
the normal state paramagnetic moment 
$M_0 = ({{\mu}}/{{\kappa}})^2 H $,   and 
${\kappa}=B_0/\pi k_{\rm B}T_{\rm c}\sqrt{8\pi N_0}$ . 
We set the Ginzburg-Landau parameter $\kappa=30$ as   
typical type-II superconductors.  

The calculations of Eqs. (\ref{eq:Eil})-(\ref{eq:scM}) 
in the vortex lattice state 
are alternatively iterated, and we obtain selfconsistent solutions
of the pair potential $\Delta({\bf r})$, vector potential ${\bf A}({\bf r})$,  
and quasi-classical Green's functions $g$, $f$ and 
$f^\dagger$.~\cite{Klein1987,IchiokaD1,IchiokaPara,Miranovic,IchiokaIntech}
We perform calculations for a scattering parameter 
$1/\tau_0=0.1$ in the Born limit and in the unitary limit, 
in addition to the clean limit $1/\tau_0=0$, 
to examine the $T$-dependence and $H$-dependences in each case. 
To calculate the paramagnetic susceptibility, 
we set paramagnetic parameter as $\mu=0.01$. 
The contributions of the paramagnetic pair-breaking are negligible 
for this very small $\mu$.  
We report the cases of 
triangular vortex lattice,
and add some results on the square vortex lattice cases 
at higher fields in the $d_{x^2-y^2}$-wave pairing. 

We note that the selfconsistent calculation of $\Delta({\bf r})$ is 
necessary to correctly estimate the $H$- and $T$- dependences of 
the vortex core size and the pair-potential's amplitude. 
For the quantitative estimate of physical quantities 
in the vortex state, we have to exactly estimate the vortex core structure, 
including the influences of the core contributions 
toward the outside of vortices. 
In the non-selfconsistent calculations, these $H$- and $T$-dependences 
are given as assumptions. 
While the calculation method of Doppler shift neglects  
the vortex core contribution, 
the vortex core gives significant contribution to 
the zero-energy DOS, as shown in Fig.1 of Ref. \onlinecite{Nakai}. 
Also in the study of two-band superconductors, 
we see the difference in the $H$-dependence of zero-energy DOS 
between the calculation of the Doppler shift methods~\cite{Bang} and 
the self-consistent Eilenberger calculation~\cite{IchiokaMgB2} 
in the clean limit. 
Therefore, the selfconsistent calculation is valuable 
for the quantitative study of properties of vortex state 
in the whole range of $H$ and $T$.  

\section{$s$-wave pairing}
\label{sec:s}
\subsection{Clean limit}

In this section, we study the spatial structure of 
the Knight shift $M_{\rm para}({\bf r})$ and the internal field distribution 
$B({\bf r})$ in the $s$-wave pairing, to estimate 
the resonance line shapes $P(M)$ and $P(B)$. 
First, we discuss behaviors in the clean limit. 
By the selfconsistent calculations, we obtain 
$M_{\rm para}({\bf r})$ and $B({\bf r})$ shown in Fig. \ref{fig1}. 

As for the $T$-dependence presented 
in Figs. \ref{fig1}(a) and \ref{fig1}(b), 
$M_{\rm para}({\bf r})$ is uniform near $T=T_{\rm c}$. 
On lowering temperature, 
$M_{\rm para}({\bf r})$ decreases outside of vortex core, 
and increases inside the vortex core. 
We see rapid increases at the vortex center at low $T$.   
Both at low $H=0.02$ and higher $H=0.1$, 
the main distribution is restricted inside the vortex core, 
$r \le \xi_0$.  
This indicates that 
the characteristic length of $M_{\rm para}({\bf r})$-distribution 
is the superconducting coherence length $\xi_0$. 
In the spatial structure of $M_{\rm para}({\bf r})$ at $H=0.02$ 
in the insets of Fig. \ref{fig1}(a), 
outside of the vortex core, 
$M_{\rm para}({\bf r})$ has flat distribution  
and $M_{\rm para}({\bf r})\sim 0$ at low $T$ and low $H$. 
At a higher field $H=0.1$ shown in the inset of Fig. \ref{fig1}(b), 
since foot of $M_{\rm para}({\bf r})$-distribution around the vortex cores
overlap each other with those of neighbor vortex cores, 
$M_{\rm para}({\bf r})$ has the spatial variation 
even outside of the vortex core. 

Also in the $T$-dependence of $B({\bf r})$ 
in Figs. \ref{fig1}(c) and \ref{fig1}(d), 
$B({\bf r})$ is uniform near $T=T_{\rm c}$. 
On lowering $T$, $B({\bf r})$ is enhanced around vortex core, 
and suppressed in the outer region. 
The difference from $M_{\rm para}({\bf r})$ is that 
the characteristic length of $B({\bf r})$ 
is the penetration depth $\lambda$. 
Therefore $B({\bf r})$ decreases monotonically 
as a function of radius $r$ from the vortex center 
until outside of vortex cores. 
In the $T$-dependence, 
increase of $B({\bf r})$ on lowering $T$ is 
not restricted in the vortex core region, 
which is determined by the inter-vortex distance rather 
than the coherence length, 
as shown in Figs. \ref{fig1}(c) and \ref{fig1}(d). 
Outside of the vortex, 
we see the structure of saddle points at midpoints between 
nearest neighbor vortices, and minimum 
at equidistant points from adjacent three vortices 
in the insets of Figs. \ref{fig1}(c) and \ref{fig1}(d).

\begin{figure}
\begin{center}
\includegraphics[width=9.5cm]{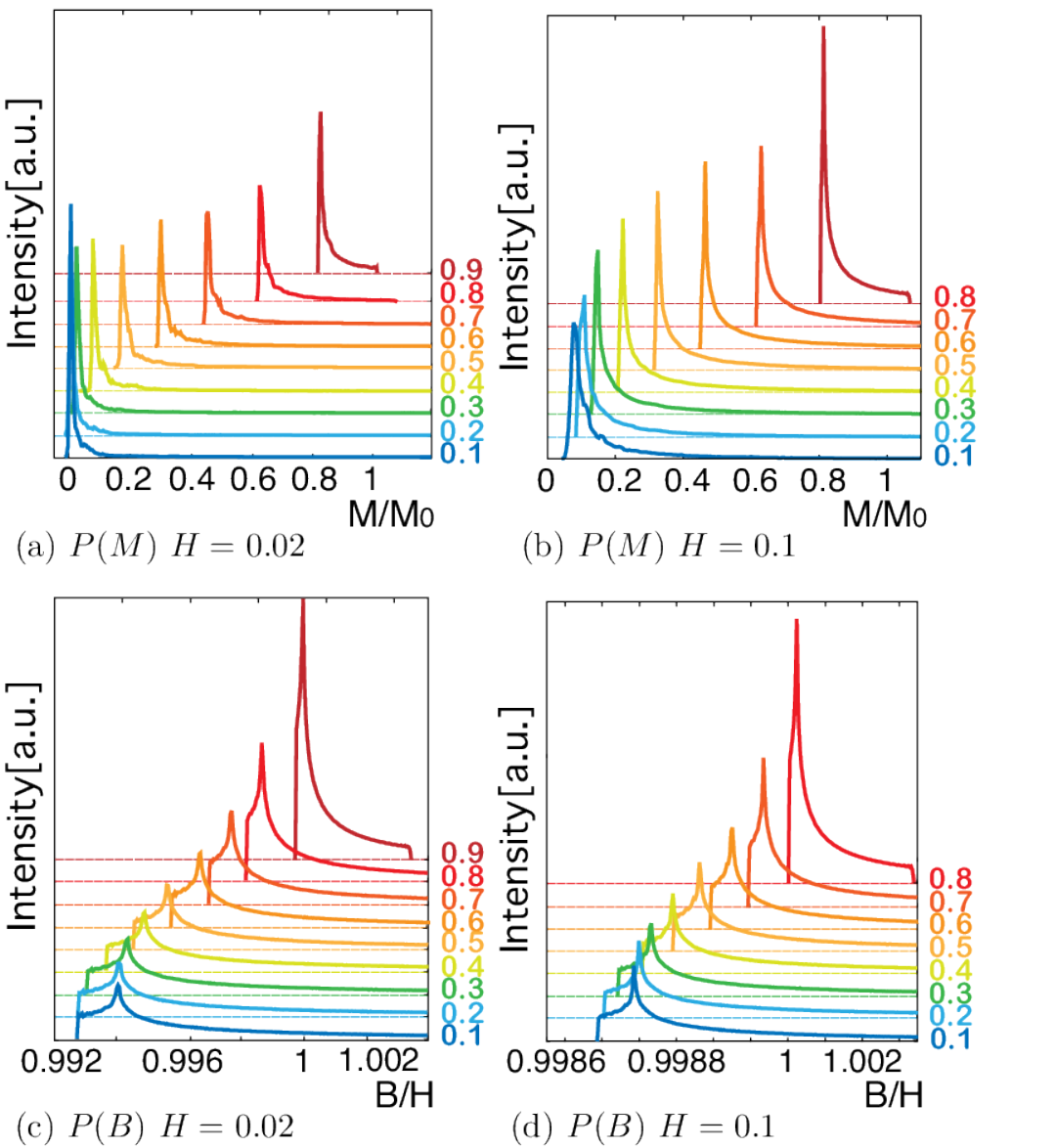}
\end{center}
\vspace{-0.7cm}
\caption{\label{fig2}
(Color online) 
Changes of the NMR resonance line shape on lowering $T$ 
in the $s$-wave pairing and in the clean limit.  
We show the Knight shift spectrum $P(M)$ 
as a function of $M/M_0$ 
for (a) $H=0.02$ and (b) $H=0.1$   
at $T/T_{\rm c}=0.1, \ 0.2, \cdots,\ 0.9$. 
For the comparison we also show 
the Redfield pattern $P(B)$ 
as a function of $B/H$ 
for (c) $H=0.02$ and (d) $H=0.1$. 
The horizontal base line for each spectrum is shifted by $T/T_{\rm c}$.  
}
\end{figure}

The above-mentioned properties of $M_{\rm para}({\bf r})$ and $B({\bf r})$ 
induce differences of the resonance line shapes of the Knight shift $P(M)$ 
and the Redfield pattern $P(B)$. 
In $P(M)$ in Figs. \ref{fig2}(a) and \ref{fig2}(b),  
the minimum edge $M_{\rm min}$ decreases on lowering $T$. 
The distribution $P(M)$ has sharp peak, and peak position $M_{\rm peak}$ 
is located near $M_{\rm min}$ in the distribution. 
This is because the peak comes from the uniform distribution outside 
of the vortex core. 
Compared with Fig. \ref{fig2}(b) at a higher field $H=0.1$, 
the peak position $M_{\rm peak}$ in $P(M)$ is shifted 
to lower $M$, and reduces to $M=0$,  
in Fig. \ref{fig2}(a) at a lower field $H=0.02$.  

Also in the Redfield pattern of $P(B)$, 
the minimum edge $B_{\rm min}$ decreases on lowering $T$. 
Difference between $P(M)$ and $P(B)$ is that 
the peak position $B_{\rm peak}$ is located at a different position from 
the minimum field $B_{\rm min}$, 
as presented in Figs. \ref{fig2}(c) and \ref{fig2}(d). 
This is because $B({\bf r})$ has the spatial distribution 
even outside of vortex core. 
That is, $B({\bf r})$ has different values 
for $B_{\rm peak}$ at the saddle point 
and for $B_{\rm min}$ 
at equidistant points from adjacent three vortices.  

\begin{figure}
\begin{center}
\includegraphics[width=8.8cm]{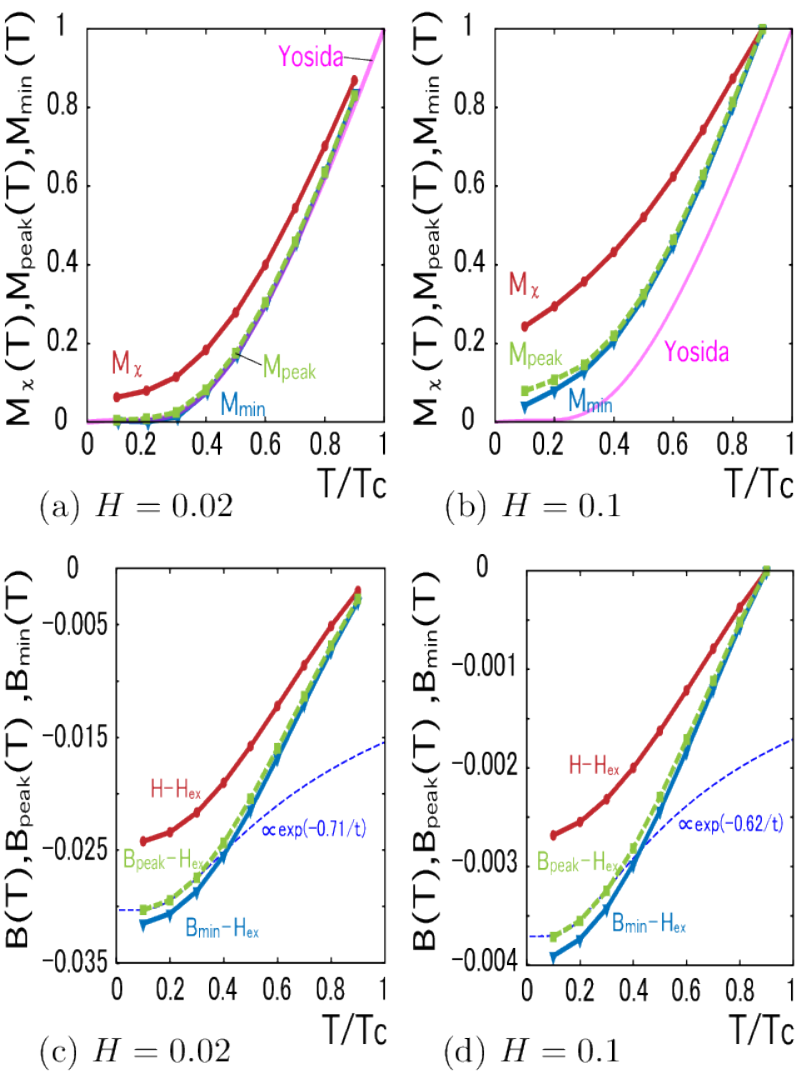}
\end{center}
\vspace{-0.7cm}
\caption{\label{fig3}
(Color online) 
(a) 
$T$-dependence of the peak position $M_{\rm peak}$, 
minimum edge $M_{\rm min}$, and 
the weighted center $M_\chi$ of the distribution $P(M)$ 
at $H=0.02$.  
We also show the $T$-dependence of the Yosida function.  
(b) 
The same as (a), but at $H=0.1$. 
(c) 
$T$-dependence of the peak position $B_{\rm peak}$ 
and the minimum field $B_{\rm min}$ of the distribution $P(B)$ at $H=0.02$. 
We plot the shift from the external field as  
$(B_{\rm peak}-H_{\rm ex})/H$, $(B_{\rm min}-H_{\rm ex})/H$, respectively.  
We also show the shift of the averaged internal field $(H -H_{\rm ex})/H$, 
which indicates the $T$-dependence of the magnetization. 
The dashed line indicates a fitting by an exponential function.
(d) 
The same as (c), but at $H=0.1$.  
These are for the $s$-wave pairing in the clean limit.  
}
\end{figure}

To discuss the $T$-dependence of $P(M)$, 
we focus on behaviors of the peak position $M_{\rm peak}$, 
the minimum edge $M_{\rm min}$, and 
the weighted center $M_\chi$ of $P(M)$. 
$M_{\chi}$ is a paramagnetic susceptibility obtained 
by the spatial average of $M_{\rm para}({\bf r})$. 
We present the $T$-dependence of $M_{\rm peak}$, $M_{\rm min}$, 
and $M_\chi$ in Figs. \ref{fig3}(a) and \ref{fig3}(b).
We also show the $T$-dependence of the Yosida function,~\cite{Yosida} 
which is for uniform states without vortices. 
At a low field $H=0.02$ in Fig. \ref{fig3}(a), 
$M_{\rm peak}(\sim M_{\rm min})$ 
shows an exponential $T$-dependence, and it coincides  
with that of the Yosida function, even in the vortex state. 
This indicates that $M_{\rm peak}$ reflects 
the local electronic structure outside of vortex cores, 
and that the exponential $T$-dependence of the $s$-wave pairing 
can be observed by $M_{\rm peak}$ even in the vortex state at low $H$.  
The paramagnetic susceptibility $M_\chi$ is larger than $M_{\rm peak}$, 
and the $T$-dependence of $M_\chi$ is a power-law,  
because it includes low energy excitations in the vortex core. 
At a higher field $H=0.1$ in Fig. \ref{fig3}(b), 
the $T$-dependence of $M_{\rm peak}$ deviates from 
that of the Yosida function, and shows a power-law $T$-dependence. 
This is because the contributions of low energy excitations 
at the vortex core extends to the outside region between vortices. 

The $T$-dependence of the peak position $B_{\rm peak}$ 
and the lower-edge $B_{\rm min}$ of the Redfield pattern $P(B)$ 
is presented in Figs. \ref{fig3}(c) and \ref{fig3}(d),  
where we show the shift from the applied external field $H_{\rm ex}$. 
From the selfconsistent solutions, we obtain $H_{\rm ex}$ as 
\begin{eqnarray} && 
H_{\rm ex}=
H
+\left\langle \left( B({\bf r})-H \right)^2\right\rangle_{\bf r}
/H 
\nonumber \\ &&   
+\frac{T}{{\kappa}^2 H} \sum_{\omega_n >0} \langle \langle  
{\rm Re} \{ 
\frac{(f^\dagger \Delta\phi
+f \Delta^\ast \phi^\ast )g}{2(g+1)} 
+\omega_n ( g-1 ) \} 
\rangle_{\bf p}\rangle_{\bf r}, 
\nonumber \\ && 
\label{eq:Mag}
\end{eqnarray} 
which is derived by Doria-Gubernatis-Rainer scaling.~\cite{WatanabeKita,Doria} 
$\langle\cdots\rangle_{\bf r}$ indicates spatial average.
The shift of the weighted center $H-H_{\rm ex}$ of $P(B)$ 
indicates the $T$-dependence of the magnetization.  
We see $B_{\rm min} < B_{\rm peak}$ until higher $T$ in these figures. 
Compared with those of Fig. \ref{fig3}(d), 
the $T$-dependence becomes weak at low $T$ in the $s$-wave pairing 
at a low field in Fig. \ref{fig3}(c). 
We also show a fitting by an exponential function for the behavior 
in the figure.

\subsection{Influence of impurity scattering} 
 
\begin{figure}
\begin{center}
\includegraphics[width=8.8cm]{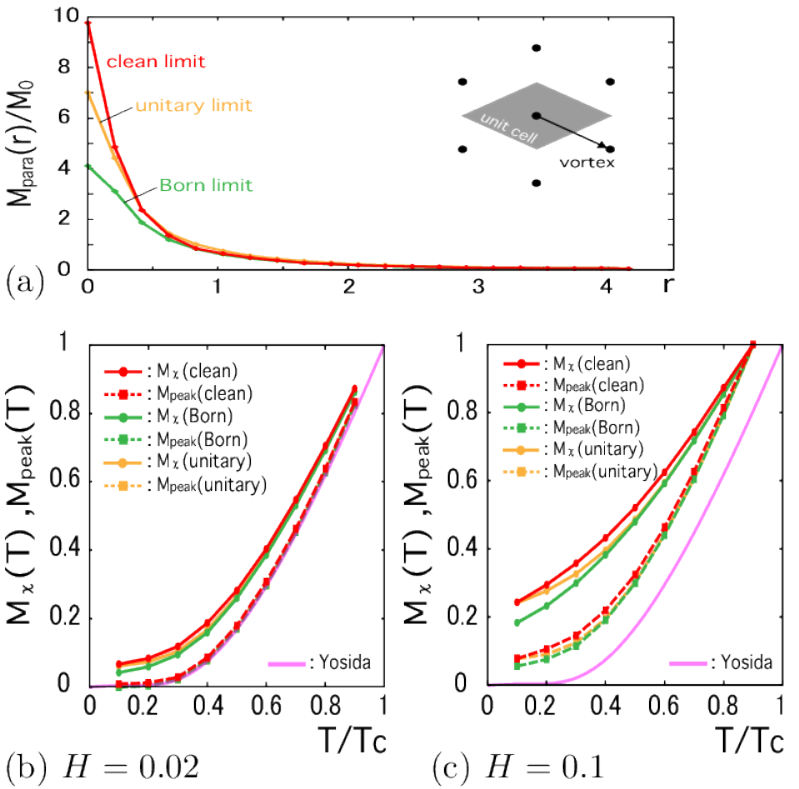}
\end{center}
\vspace{-0.7cm}
\caption{\label{fig4}
(Color online) 
(a) Profile of $M_{\rm para}({\bf r})$ as a function of radius $r/\xi_0$ 
from the vortex center along the nearest neighbor vortex direction 
at $T/T_{\rm c}=0.1$ and $H=0.1$ for the $s$-wave pairing.  
We show the cases of the Born limit and the unitary limit 
of $1/\tau=0.1$, with that of the clean limit. 
(b) 
$T$-dependence of the peak position $M_{\rm peak}$ and 
the weighted center $M_\chi$ of the distribution $P(M)$ 
at $H=0.02$ for the $s$-wave pairing 
in the Born limit and the unitary limit of $1/\tau=0.1$ 
in addition to the clean limit case. 
We also show the $T$-dependence of the Yosida function.  
(c) 
The same as (b), but at $H=0.1$. 
}
\end{figure}

To discuss influences of the impurity scatterings in the vortex state for  
the $s$-wave pairing, we show the profile of $M_{\rm para}({\bf r})$ 
in Fig. \ref{fig4}(a). 
At the vortex core,  $M_{\rm para}({\bf r})$ is suppressed by the 
impurity scatterings. 
The suppression of $M_{\rm para}({\bf r})$ is stronger in the Born limit, 
compared with the case of the unitary limit. 
This comes from the fact that 
low energy states at the vortex core 
is smaller in the Born limit than in the unitary 
limit.\cite{eshrig2002} 
On the other hand, 
at the outside region of the vortex core $M_{\rm para}({\bf r})$ 
is not changed by the impurity scattering. 
This indicates that the non-magnetic impurity scattering does 
not break the $s$-wave superconductivity in the uniform state, 
which is similar situation as in Anderson's theorem 
at a zero field.\cite{Anderson,AbrikosovGorkov}

In Figs. \ref{fig4}(b) and \ref{fig4}(c), we present the $T$-dependence of 
$M_{\rm peak}$ and $M_{\chi}$ in the presence of the impurity scattering. 
The behavior of $M_{\rm peak}$ whose contributions are 
from outside of the vortex core 
is not changed by the non-magnetic impurities. 
In the $T$-dependence of $M_\chi$ which includes contributions 
of the vortex cores, 
there are small changes by the impurity scattering at low $T$. 
The changes are larger at higher $H$ in Fig. \ref{fig4}(c). 

\subsection{Magnetic field dependence}

\begin{figure}
\begin{center}
\includegraphics[width=7.5cm]{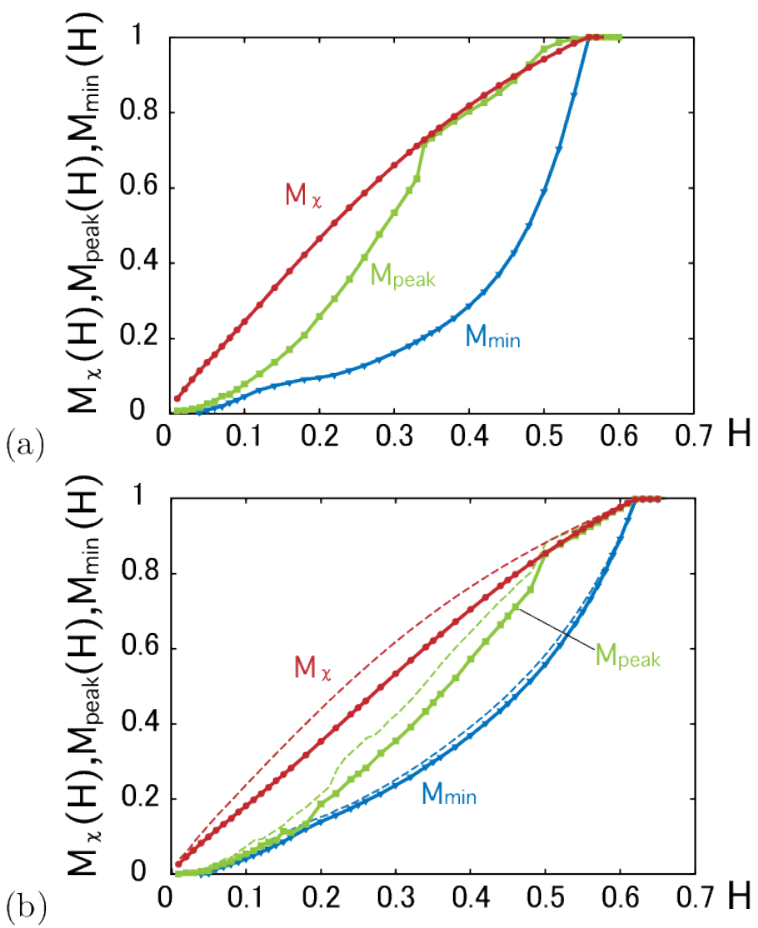}
\end{center}
\vspace{-0.7cm}
\caption{\label{fig5}
(Color online) 
(a) 
$H$-dependence of the peak position $M_{\rm peak}$, 
the minimum edge $M_{\rm min}$, and 
the weighted center $M_\chi$ of the distribution $P(M)$ 
in the clean limit at $T/T_{\rm c}=0.1$ 
for the $s$-wave pairing. 
(b) 
The same as (a), but 
in the Born limit (solid lines) and 
in the unitary limit (dashed lines) of $1/\tau=0.1$.  
}
\end{figure}

\begin{figure}
\begin{center}
\includegraphics[width=5.5cm]{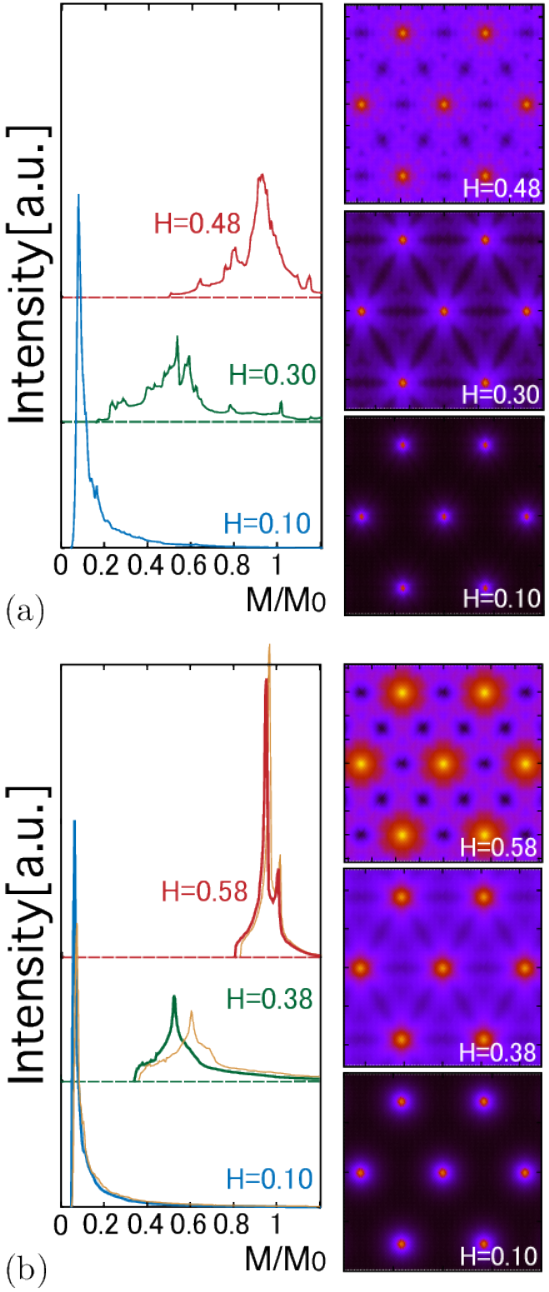}
\end{center}
\vspace{-0.7cm}
\caption{\label{fig6}
(Color online) 
(a) Resonance line shape of $P(M)$ [left panels] and 
density plots of $M_{\rm para}({\bf r})$ [right panels] 
at $H=0.10$, 0.30, and 0.48 in the clean limit 
for the $s$-wave pairing. 
$T/T_{\rm c}=0.1$. 
The horizontal base line for each $P(M)$ is shifted. 
(b) The same as (a), but 
at $H=0.10$, 0.38, and 0.58 in the Born limit with $1/\tau=0.1$. 
We also show $P(M)$ for the unitary limit by thin lines 
in the left panel. 
}
\end{figure}

Figure \ref{fig5} presents the $H$-dependence of $M_{\rm peak}$, 
$M_{\rm min}$, and $M_\chi$ in the $s$-wave pairing.   
At low $T$, the paramagnetic susceptibility $M_\chi$ is 
proportional to the zero-energy DOS. 
In Fig. \ref{fig5}, we see the linear $H$-dependence, 
$M_\chi \propto H$, at low $H$ both in the clean limit and 
in the presence of the impurity scatterings.  
However, since $M_{\rm peak} < M_\chi$ at low fields,
$M_{\rm peak}$ shows different $H$-dependence from the linear relation. 
On the other hand, $M_{\rm peak} \sim M_\chi$ at higher fields.  
These behaviors are related to the line shape of $P(M)$ and 
the  spatial structure of $M_{\rm para}({\bf r})$, 
as presented in Fig. \ref{fig6}.
At a low field $H=0.1$, $M_{\rm para}({\bf r})$ is localized 
within the vortex core, and $P(M)$ has a sharp peak 
at the minimum edge $M_{\rm min}$. 
Thus, $M_{\rm min} \sim M_{\rm peak} < M_\chi$. 
At higher fields, the main distributions of $M_{\rm para}({\bf r})$ 
are connected by the tails between neighbor vortices. 
Thus, the structures of saddle points and minimum points appear 
in the outside region of the vortex core. 
Therefore, the peak position of $P(M)$, coming from the saddle points, 
moves to larger-$M$ position from the minimum-edge $M_{\rm min}$ 
in the distribution $P(M)$. 
Therefore, $M_{\rm min}<M_{\rm peak} \sim M_\chi$ at higher fields.  

In the clean limit in Fig. \ref{fig6}(a), 
since the inter-vortex connection of $M_{\rm para}({\bf r})$ 
has fine structures, the resonance line shape of $P(M)$ has fine structure 
with many sub-peaks. 
In the presence of the impurity scattering, 
as presented in Fig. \ref{fig6}(b), 
the inter-vortex connection of $M_{\rm para}({\bf r})$ are smeared. 
Thus, the fine structures of $P(M)$ is smeared to smooth spectrum shape.


\section{$d_{x^2-y^2}$-wave pairing} 
\label{sec:d}

\subsection{Clean limit}

In unconventional superconductors, 
the anisotropic pairing function changes the sign on the Fermi surface.  
And due to the node structure of the pairing function, there appear 
low energy states within the superconducting gap. 
As an example of the anisotropic superconductivity, 
we study the case of $d_{x^2-y^2}$-wave pairing, and discuss 
how behaviors of the NMR resonance line shape change 
from the case of $s$-wave pairing in the previous section. 
 
\begin{figure}
\begin{center}
\includegraphics[width=9.5cm]{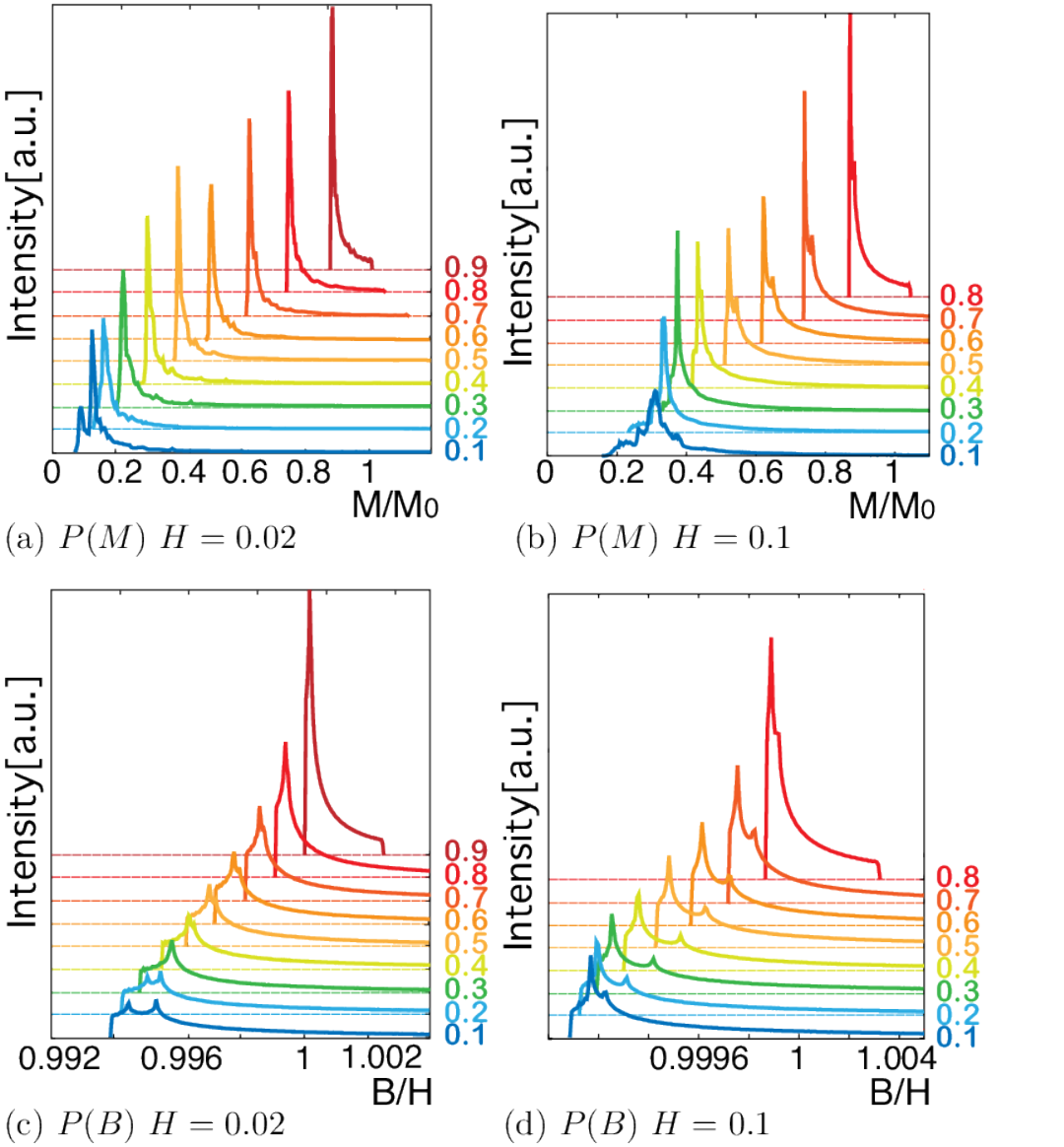}
\end{center}
\vspace{-0.7cm}
\caption{\label{fig7}
(Color online) 
Change of the NMR resonance line shape on lowering $T$ 
in the $d_{x^2-y^2}$-wave pairing and in the clean limit.  
We show the Knight shift spectrum $P(M)$ for    
(a) $H=0.02$ and (b) $H=0.1$   
at $T/T_{\rm c}=0.1, \ 0.2, \cdots,\ 0.9$. 
For the comparison we also show 
the Redfield pattern spectrum $P(B)$ for  
(c) $H=0.02$ and (d) $H=0.1$.  
The horizontal base line for each spectrum is shifted by $T/T_{\rm c}$.  
}
\end{figure}

\begin{figure}
\begin{center}
\includegraphics[width=8.8cm]{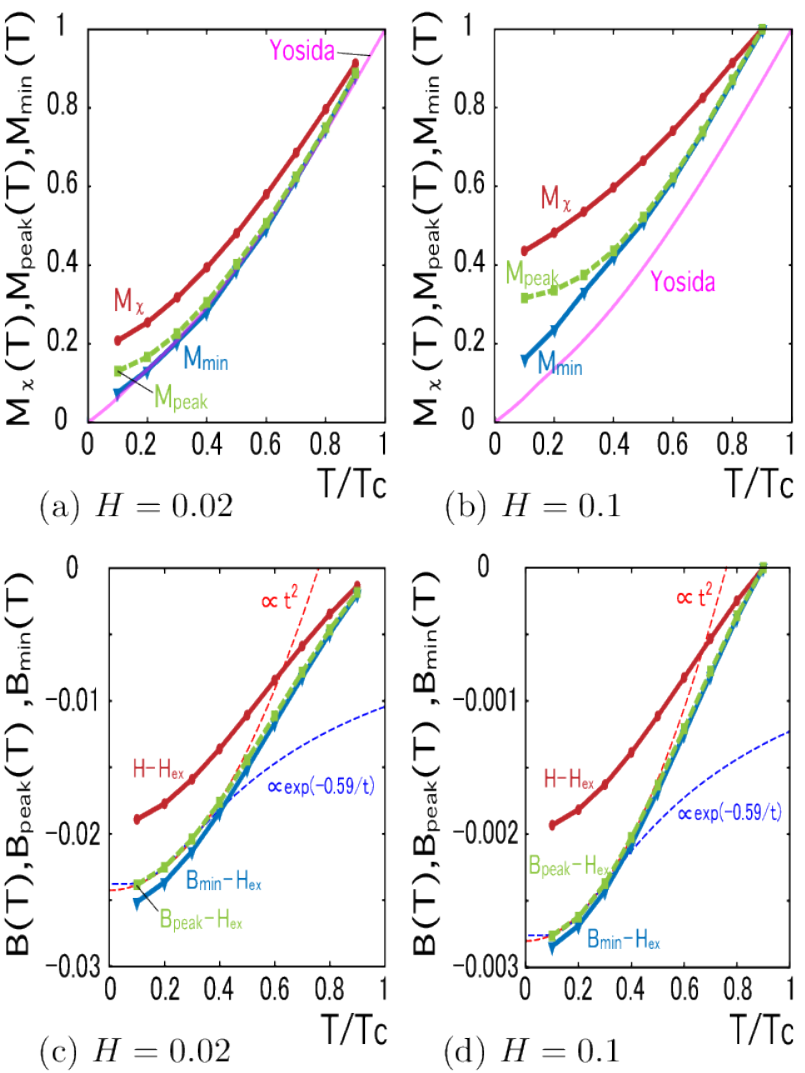}
\end{center}
\vspace{-0.7cm}
\caption{\label{fig8}
(Color online) 
(a) 
$T$-dependence of the peak position $M_{\rm peak}$, 
minimum edge $M_{\rm min}$, and 
the weighted center $M_\chi$ of the distribution $P(M)$ 
at $H=0.02$.  
We also show the $T$-dependence of the Yosida function of 
the $d_{x^2-y^2}$-wave pairing.  
(b) 
The same as (a), but at $H=0.1$. 
(c) 
$T$-dependence of the peak position $B_{\rm peak}$ 
and the minimum field $B_{\rm min}$ of the distribution $P(B)$ at $H=0.02$. 
We plot the shift from the external field as  
$(B_{\rm peak}-H_{\rm ex})/H$, $(B_{\rm min}-H_{\rm ex})/H$, respectively.  
We also show the shift of the averaged internal field $(H -H_{\rm ex})/H$, 
which indicates the $T$-dependence of the magnetization. 
Dashed lines indicate fittings by a exponential function and a power function. 
(d) 
The same as (c), but at $H=0.1$.  
These are for the $d_{x^2-y^2}$-wave pairing in the clean limit.  
}
\end{figure}

In Fig. \ref{fig7}, we present the temperature evolution of 
the NMR resonance line shape $P(M)$ and $P(B)$ 
in the $d_{x^2-y^2}$-wave pairing at $H=0.02$ and 0.1.  
In the Knight shift spectrum $P(M)$ in Figs. \ref{fig7}(a) and \ref{fig7}(b), 
at higher $T>0.4T_{\rm c}$,   
the peak position $M_{\rm peak}$ is located at the minimum edge $M_{\rm min}$, 
as in the $s$-wave pairing. 
However, at lower $T$, position of $M_{\rm peak}$ deviates from $M_{\rm min}$. 
The $T$-dependences of $M_{\rm peak}$, $M_{\rm min}$, 
and the weighted center $M_\chi$ 
are presented in Figs. \ref{fig8}(a) and \ref{fig8}(b). 
Due to the low energy excitations by the node of the pairing function, 
the $T$-dependence is different from that in the $s$-wave pairing, 
including the $T$-dependence of the Yosida function for a uniform state 
in the $d_{x^2-y^2}$-wave pairing. 
At a low field $H=0.02$, $M_{\rm peak}$ follow 
the $T$-dependence of the Yosida function at higher $T>0.4T_{\rm c}$, 
but deviates from it at lower $T$. 
$M_{\rm min}$ follows the power-law $T$-dependence of the Yosida function 
until low $T$. 
The $T$-dependence of the weighted center $M_\chi$ also shows the power law 
behavior as a function of $T$, and $M_\chi > M_{\rm peak}$.  

The Redfield pattern $P(B)$ is presented 
in Figs. \ref{fig7}(c) and \ref{fig7}(d). 
In the $d_{x^2-y^2}$-wave pairing, we see the second peak in $P(B)$. 
It comes from the fourfold vortex core shape 
in the $d_{x^2-y^2}$-wave pairing.\cite{IchiokaD1,Ichioka1996}  
Compared to the $s$-wave pairing case 
in Figs. \ref{fig2}(c) and Figs. \ref{fig2}(d), 
the peak position $B_{\rm peak}$ and the minimum edge $B_{\rm min}$ 
are larger in the $d_{x^2-y^2}$-wave pairing case 
in Figs. \ref{fig7}(c) and \ref{fig7}(d). 
Even at low $T$ ($T/T_{\rm c} \le 0.2$), 
$B_{\rm peak}$ and $B_{\rm min}$ continue to decrease on lowering $T$ 
in the $d_{x^2-y^2}$-wave pairing. 
These behaviors are also seen in Figs. \ref{fig8}(a) and \ref{fig8}(b), 
where the low $T$ behaviors are fitted by $T^2$-function. 
They are related to the difference of the $T$-dependence of 
the superfluid density between the $s$-wave pairing and the $d_{x^2-y^2}$-wave pairing. 
This is because 
the internal field $B({\bf r})$ determined by Eq. (\ref{eq:scH}) 
and the magnetization calculated by Eq. (\ref{eq:Mag}) have a term 
with a factor  $\kappa^{-2} \propto \lambda^{-2}$, 
which is proportional to the superfluid density.     

\subsection{Influence of impurity scattering} 
 
In Eilenberger Eq. (\ref{eq:Eil}), 
the Fermi surface average $\langle f \rangle_{\bf k}$ of the impurity scattering 
is canceled by the sign change of the pairing function on the Fermi surface. 
Therefore, in the $d_{x^2-y^2}$-wave pairing, 
the influence of the impurity scattering is different 
from the $s$-wave pairing.  
For example, 
non-magnetic impurity scattering suppresses 
the superconducting transition temperature $T_{\rm c}$ 
in the $d_{x^2-y^2}$-wave pairing.

\begin{figure}
\begin{center}
\includegraphics[width=8.8cm]{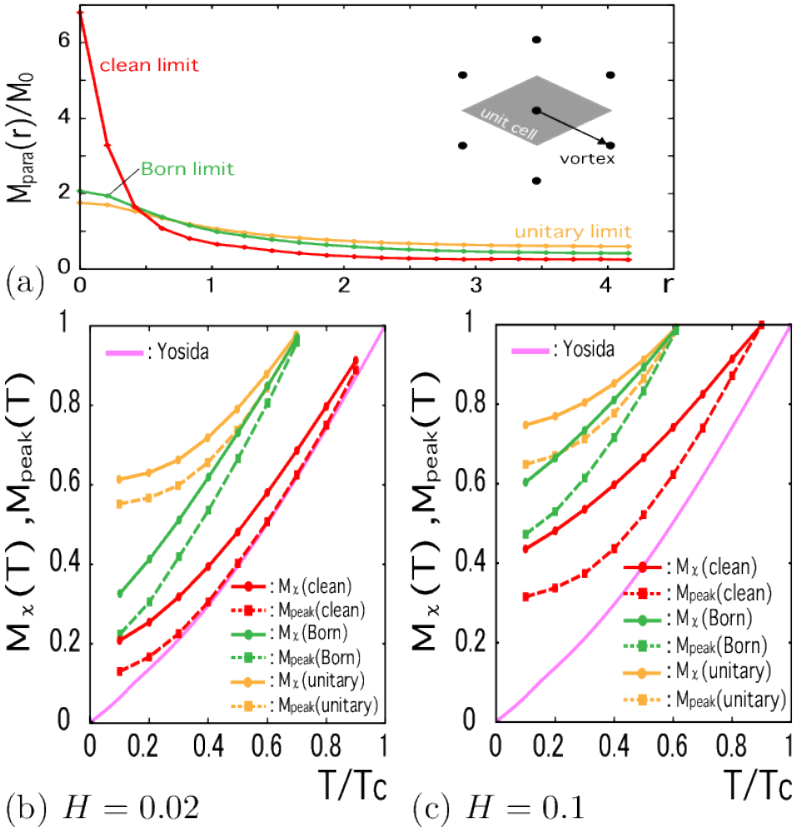}
\end{center}
\vspace{-0.7cm}
\caption{\label{fig9}
(Color online) 
(a) Profile of $M_{\rm para}({\bf r})$ as a function of radius $r/\xi_0$ 
from the vortex center along the nearest neighbor vortex direction
at $T/T_{\rm c}=0.1$ and $H=0.1$ 
for the $d_{x^2-y^2}$-wave pairing.  
We show the cases of the Born limit and the unitary limit 
of $1/\tau=0.1$, with that of the clean limit. 
(b) 
$T$-dependence of the peak position $M_{\rm peak}$ (dashed lines) and 
the weighted center $M_\chi$ (solid lines) 
of the distribution $P(M)$ at $H=0.02$  
in the Born limit and the unitary limit of $1/\tau=0.1$.  
We also show those of the clean limit, 
and the $T$-dependence of the Yosida function for the $d_{x^2-y^2}$-wave pairing.   
(c) 
The same as (b), but at $H=0.1$. 
}
\end{figure}

In Fig. \ref{fig9}(a), we present profiles of $M_{\rm para}({\bf r})$ 
around a vortex with and without non-magnetic impurity scattering. 
At the vortex center, height of $M_{\rm para}({\bf r})$ is suppressed 
by the impurity  scattering.  
Outside of the vortex core,  
$M_{\rm para}({\bf r})$ is enhanced toward the recovery  
to the normal state value.  
These effect is stronger in the unitary limit than in the  Born limit. 

In Figs. \ref{fig9}(b) and  \ref{fig9}(c), 
we show the $T$-dependence of $M_{\rm peak}$ and $M_\chi$ in the presence of 
impurity scattering. 
Compared with the case of the clean limit, 
both $M_{\rm peak}$ and $M_\chi$ shift to higher $M$ 
by the impurity scattering, because
the superconducting transition temperature is suppressed. 
Values of $M_{\rm peak}$ and $M_\chi$ are larger 
in the unitary limit than in the Born limit, 
because the low energy states by the impurity scattering 
are more enhanced in the unitary limit. 
Both at $H=0.02$ and $H=0.1$, 
we find $M_\chi > M_{\rm peak}$ also in the presence of the impurity scattering. 
In the unitary limit, the $T$-dependences are saturated, and 
$M_{\rm peak}$ and $M_\chi$ are, respectively, 
reduces to higher values at $T \rightarrow 0$. 

\subsection{Magnetic field dependence}

\begin{figure*}
\begin{center}
\includegraphics[width=18.0cm]{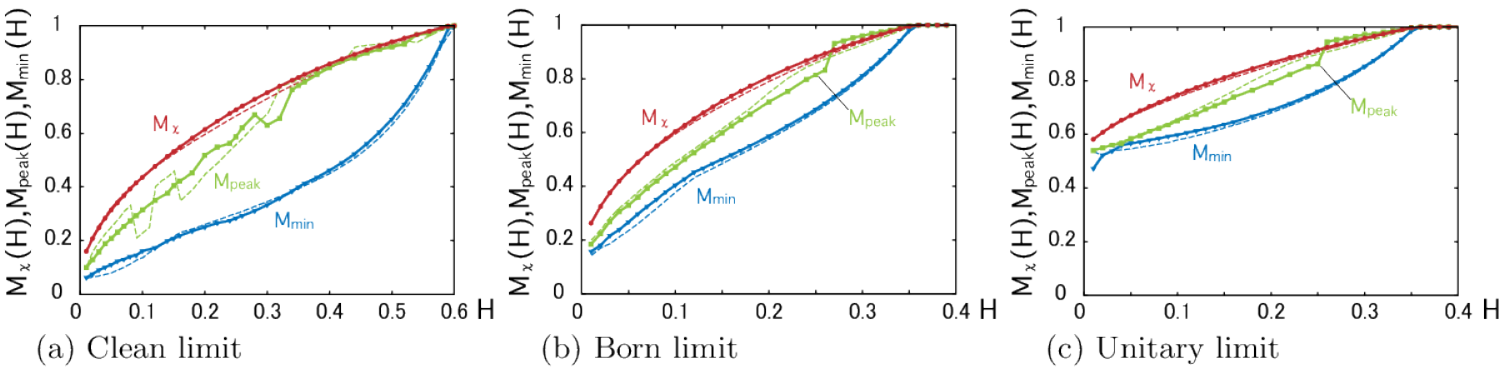}
\end{center}
\vspace{-0.7cm}
\caption{\label{fig10}
(Color online) 
(a) 
$H$-dependence of the peak position $M_{\rm peak}$, 
the minimum edge $M_{\rm min}$, and 
the weighted center $M_\chi$ of the distribution $P(M)$ 
in the clean limit at $T/T_{\rm c}=0.1$ 
for the $d_{x^2-y^2}$-wave pairing. 
The solid (dashed) lines are for the triangular (square) vortex lattice. 
(b) The same as (a), but in the Born limit of $1/\tau=0.1$.  
(c) The same as (b), but in the unitary limit. 
}
\end{figure*}
\begin{figure*}
\begin{center}
\includegraphics[width=18cm]{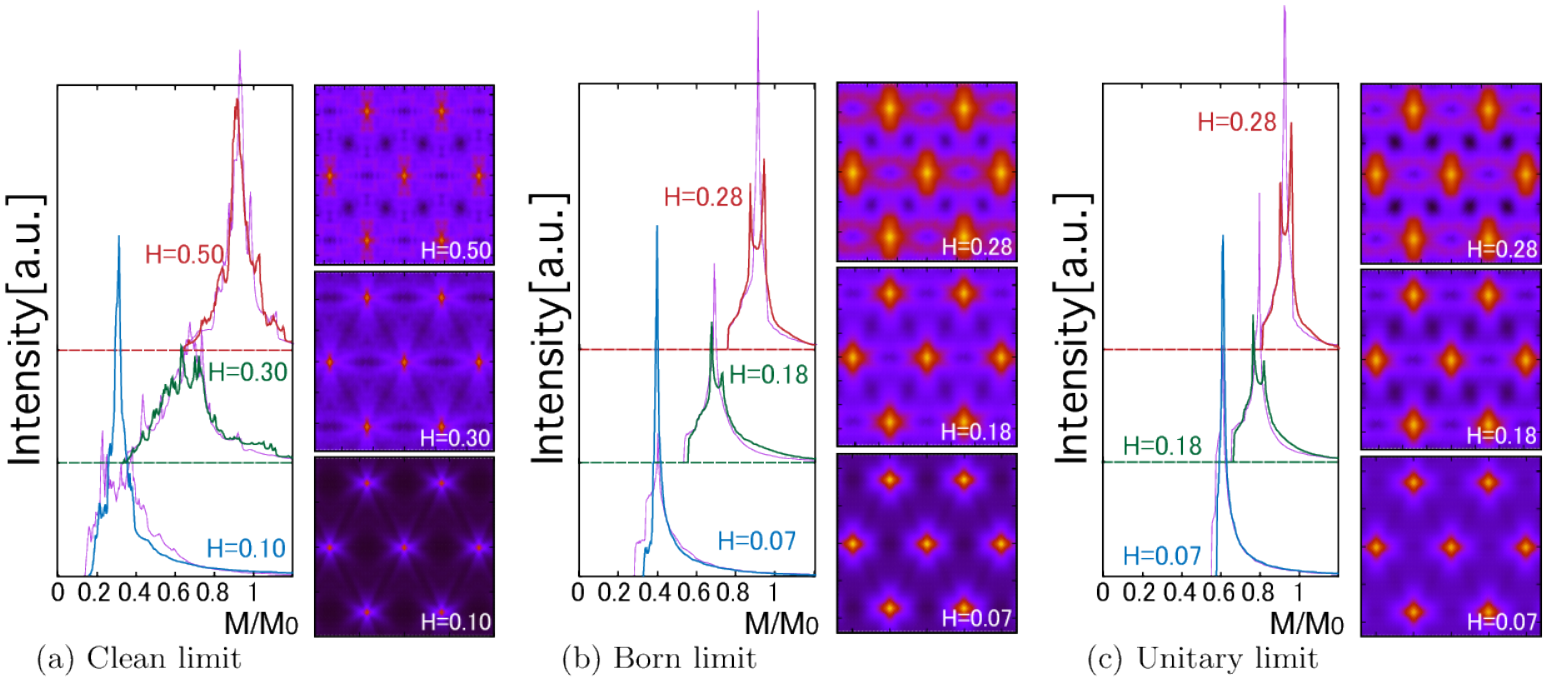}
\end{center}
\vspace{-0.7cm}
\caption{\label{fig11}
(Color online) 
(a) Line shape of $P(M)$ [left panels] and 
density plots of $M_{\rm para}({\bf r})$ [right panels] 
at $H=0.10$, 0.30, and 0.50 for triangular vortex lattice 
in the clean limit for the $d_{x^2-y^2}$-wave pairing. 
$T/T_{\rm c}=0.1$. 
We also show $P(M)$ for the square vortex lattice by thin lines 
in the left panels. 
The horizontal base line for each $P(M)$ is shifted. 
(b) The same as (a), but 
at $H=0.07$, 0.18, and 0.28 in the Born limit with $1/\tau=0.1$. 
(c) The same as (b), but in the unitary limit.
}
\end{figure*}

In Fig. \ref{fig10}, we show the $H$-dependence of $M_{\rm peak}$, 
$M_{\rm min}$, and $M_\chi$ at $T/T_{\rm c}=0.1$. 
At the low $T$, since $M_\chi$ is proportional to zero-energy DOS, 
we see the relation $M_\chi \propto \sqrt{H}$ in the low $H$ range 
due to the Volovik effect. 
By the impurity scattering, 
$H_{\rm c2}$ is suppressed by the suppression of $T_{\rm c}$.
Thus, both $M_{\rm peak}$ and $M_\chi$ shift to higher $M$, 
compared with the clean limit case. 
In the unitary limit, $M_\chi$, $M_{\rm min}$ and $M_{\rm peak}$ 
approach finite values in the limit $H \rightarrow 0$.   
In all cases with and without impurity scattering, 
$M_{\rm min} < M_{\rm peak} < M_\chi$ at the low $H$ range, 
and $M_{\rm min} < M_{\rm peak} \sim M_\chi$ 
at the high $H$ range near $H_{\rm c2}$. 

To discuss these behaviors, we present the resonance line shape $P(M)$ 
of the Knight shift and the spatial structure of $M_{\rm para}({\bf r})$ 
in Fig. \ref{fig11} in the clean limit, in the Born limit, 
and in the unitary limit.  
We present $P(M)$ also for the square vortex lattice case 
in addition to the triangular vortex lattice case, 
because the square lattice is stabilized at higher $H$ 
in the $d_{x^2-y^2}$-wave pairing.\cite{IchiokaD1} 
The following discussions do not seriously depend 
on the shape of the vortex lattice. 
In the clean limit, due to the spectrum with many sub-peaks in $P(M)$,  
the main peak position $M_{\rm peak}$ is scattered in the $H$-dependence 
in Fig. \ref{fig10}(a). 
These sub-peak structure in the clean limit is smeared 
by the impurity scattering. 
$P(M)$ in the unitary limit is shifted to higher $M$, compared to 
the Born limit case. 
The spectrum of $P(M)$ has similar shape in both limits. 
In these spectra of $P(M)$, the main peak is located 
near minimum edge $M_{\rm min}$ at low fields, and it is shifted to middle 
of the $P(M)$-distribution at higher $H$. 
These are related to the spatial structure of $M_{\rm para}({\bf r})$. 
In the $d_{x^2-y^2}$-wave pairing, zero-energy DOS at the vortex center extends outside 
towards the node direction.\cite{IchiokaD1,Ichioka1996}   
These tails of zero-energy DOS make interference with those of 
neighbor vortices, and form inter-vortex connections of $M_{\rm para}({\bf r})$. 
Therefore, we see saddle points and minimum points at the boundary region 
of a unit cell of the vortex lattice. 
This is a reason why 
the peak position $M_{\rm peak}$ by the contribution of the saddle points 
are deviated from the minimum $M_{\rm min}$. 
The fine structure of the inter-vortex connection of 
$M_{\rm para}({\bf r})$ is smeared by the impurity scattering. 
By the smearing, 
$P(M)$ becomes smooth spectrum shape 
as seen in Figs. \ref{fig11}(b) and \ref{fig11}(c).

\section{Summary} 
\label{sec:summary}

We studied the resonance line shape of the NMR spectrum in the vortex states 
based on quantitative calculation by Eilenberger theory, 
to clarify the difference of Knight shift spectrum $P(M)$ and 
the Redfield pattern spectrum $P(B)$. 
The former is the case when the hyperfine coupling constant $A_{\rm hf}$ 
is large, and the latter is the opposite case of negligible $A_{\rm hf}$. 
Since the characteristic length for the spacial structure of 
the paramagnetic moment $M_{\rm para}({\bf r})$ is the coherence length, 
dominant distribution of $M_{\rm para}({\bf r})$ is restricted within 
the vortex core region, 
and in the outside region $M_{\rm para}({\bf r})$ is uniform with 
minimum value $M_{\rm min}$. 
Thus, the peak of $P(M)$ comes from the signal 
outside of vortex core, and the peak position $M_{\rm peak}$ is 
located near the minimum edge $M_{\rm min}$ of $P(M)$ at low fields. 
On the other hand, 
the characteristic length for the spacial structure of 
the internal magnetic field $B({\bf r})$ is the penetration length, 
spatial variation of $B({\bf r})$ occurs even outside of the vortex core.  
As $B({\bf r})$ has different values for $B_{\rm peak}$ at the saddle points 
and for $B_{\rm min}$ at the minimum points, 
the peak position $B_{\rm peak}$ is apart from the 
minimum edge $B_{\rm min}$ in the Redfield pattern $P(B)$.  

We estimated the temperature dependence and the magnetic field dependence of 
the Knight shift spectrum $P(M)$, and studied the differences 
between the full gap $s$-wave pairing case and the anisotropic 
$d_{x^2-y^2}$-wave pairing case. 
In addition to results in the clean limit, 
we also discussed the influence of the impurity scattering both 
in Born limit and in the unitary limit.   
To extract the characteristic $H$-dependence of zero-energy DOS $N(E=0)$, 
we have to evaluate the weighted center $M_\chi$ of $P(M)$. 
Since $M_\chi \propto N(E=0)$, 
we expect $M_\chi \propto H$ for the $s$-wave pairing, 
and $M_\chi \propto \sqrt{H}$ for the $d_{x^2-y^2}$-wave pairing with line nodes. 
It is noted that 
the peak position $M_{\rm peak}$ of $P(M)$ deviates from $M_\chi$.  
At low fields, signal of the peak position $M_{\rm peak}$ 
can be used to 
observe the $T$-dependence of the Yosida function, 
which distinguish the pairing symmetry, even in the vortex state,  
because signal at $M_{\rm peak}$ selectively 
comes from the outside of the vortex core. 

The NMR spectrum in the multi-gap superconductors, 
such as Fe-based superconductors and ${\rm MgB_2}$, 
is one of interesting topics, and belongs to future studies. 
There, the wighted center $M_\chi$ of $P(M)$ will follow the 
characteristic $H$-dependence of zero-energy DOS reflecting 
low energy excitations in the small-gap band.~\cite{Bang,IchiokaMgB2} 
And it is also interesting to study 
the $H$-dependence of the peak position $M_{\rm peak}$, 
which will deviate from $M_\chi$. 

We hope that these theoretical estimates of $P(M)$ and $P(B)$ 
will be confirmed by the NMR experiment, 
and will be used for the analysis of the pairing symmetry and 
contributions of non-magnetic impurity scattering in the superconducting states 
by the $T$-dependence and the $H$-dependence of the NMR spectrum.



\end{document}